\documentclass[nofootinbib,superscriptaddress,a4paper,twocolumn, floatfix,aps]{revtex4-2}

\usepackage{dcolumn}% Align table columns on decimal point
\usepackage{bm}% bold math
\usepackage{graphicx}
\usepackage{float}
\usepackage{makecell}
\usepackage{subcaption}
\usepackage{amsmath}
\usepackage[ruled,vlined]{algorithm2e}
\usepackage{amsmath, amssymb} % For math symbols 
\usepackage{blindtext}
\usepackage{blindtext}
\usepackage{booktabs}

\usepackage{algpseudocode} % pseudocode
\usepackage{booktabs} % tables
\usepackage{multirow}
\usepackage{subcaption}
\usepackage[ruled,vlined]{algorithm2e}

\usepackage{braket}
\usepackage{float}
\usepackage{amsmath,amsfonts,amssymb}
\DeclareMathAlphabet{\mathbbold}{U}{bbold}{m}{n}
\usepackage{svg}

\usepackage{parskip}
\usepackage{inconsolata} % crisp monospace font
\usepackage{xcolor}

\usepackage{tikz}
\usetikzlibrary{calc}

\newcommand{\butadiene}{
\tikz[baseline=-0.5ex,scale=0.35]{
  \coordinate (A) at (0,-0.5);
  \coordinate (B) at (1,0.5);
  \coordinate (C) at (2,-0.5);
  \coordinate (D) at (3,0.5);

  \draw[line width=0.8pt] (A)--(B)--(C)--(D);

  \draw[line width=0.8pt] ($(A)+(-0.10,0.10)$)--($(B)+(-0.10,0.10)$);

  \draw[line width=0.8pt] ($(C)+(0.10,-0.10)$)--($(D)+(0.10,-0.10)$);
}
}

\newcommand{\butadieneSecond}{
\tikz[baseline=-0.5ex,scale=0.35]{
  \coordinate (A) at (0,-0.5);
  \coordinate (B) at (1.1,0.5);
  \coordinate (C) at (2.1,-0.5);
  \coordinate (D) at (3,0.5);
  \coordinate (E) at (0.9,0.5);
  \coordinate (F) at (1.9,-0.5);
  \draw[line width=0.8pt] (A)--(E)--(F);
  \draw[line width=0.8pt] (B)--(C)--(D);

  \fill ($(A)+(-0.1,0.25)$) circle (0.08);
  \fill ($(D)+(0.1,-0.25)$) circle (0.08);
}
}

\usepackage{listings}
\usepackage[T1]{fontenc}
\usepackage{textcomp} % Required if you use upquote=true in listings

\lstdefinestyle{minimalpython}{
    language=Python,
    basicstyle=\ttfamily\small,
    stringstyle=\color{green!40!black},
    keywordstyle=\bfseries\color{black},   % <- keywords now black
    commentstyle=\color{gray}\itshape,
    showstringspaces=false,
    breaklines=true,
    tabsize=4,
    numbers=none,
    frame=none,
    backgroundcolor=\color{white} % fully clean background
}
\DeclareMathOperator*{\argmax}{arg\,max}
\DeclareMathOperator*{\argmin}{arg\,min}
\providecommand{\Ng}{N_{\mathrm{g}}}
\providecolor{surH}{RGB}{31,119,180}   % surrogate hydrogen

\usepackage[dvipsnames]{xcolor}
\definecolor{unia}{HTML}{1F3A68}   
\definecolor{unia2}{HTML}{2E6DB4} 
\usepackage{hyperref}

\usepackage{hyperref}
\hypersetup{
  colorlinks=true,
  linkcolor=unia2,     % internal links (sections, equations, figures)
  citecolor=unia,     % \cite links
  urlcolor=unia,      % \url and \href
  anchorcolor=unia2
}

\makeatletter
\let\orig@startsection\@startsection
\def\@startsection#1#2#3#4#5#6{%
  \orig@startsection{#1}{#2}{#3}{#4}{#5}{#6\color{unia}}}
\makeatother

\begin{document}

\title{Alchemically Transferable Parameters for Variational Quantum Eigensolvers}

\author{Jonas Motyl}
\affiliation{{Institute for Computer Science, University of Augsburg, Germany }}
\author{Jakob~S.~Kottmann}
\email[E-mail:]{jakob.kottmann@uni-a.de}
\affiliation{{Institute for Computer Science, University of Augsburg, Germany }}
\affiliation{{Center for Advanced Analytics and Predictive Sciences, University of Augsburg, Germany }}

\date{\today}

\begin{abstract}
Machine-learned circuit parameters could remove one of the main bottlenecks of
variational quantum eigensolvers: the classical optimization loop, with its
typical barren plateaus, local minima, and costly energy and gradient evaluations, is replaced by a direct prediction. Prior
work showed that machine-learning models paired with a structured circuit design can transfer across system size and structure within
hydrogenic systems. The next step: Extending them
to organic molecules would seem to require re-training on use-case-specific training sets. In this work, we show that this is not necessary by establishing an alchemical approach. 
A geometric projection places surrogate hydrogen atoms along the bond directions
of the target molecule and so maps its local bonding environment onto small
hydrogenic fragments. 
Existing hydrogenic solvers can then be reused as they
are. 
The only new quantity is a single projection radius per bond type, which
is fitted once on a small reference molecule and reused unchanged thereafter.
We demonstrate the approach on the separable-pair circuit designs that have been used in prior work and extend the approach to multi-circuit approaches that span effective bases for quantum subspaces.
\end{abstract}

\maketitle

%\tableofcontents

%\section{\label{sec:intro}Introduction}
The electronic structure problem is widely regarded as one of the most promising applications of quantum computing.~\cite{quantum-comp-chem, bharti2022noisy, anand2022quantum, Schleich2025book} Its central objective is to determine the low-lying eigenvalues and corresponding eigenstates of the electronic Hamiltonian, providing fundamental insights into molecular properties and enabling potential applications in materials science, catalysis, and drug discovery.
Besides these static properties, direct access to the low-energy spectrum via low-depth circuits is an important building block towards dynamical properties.~\cite{schleich2025, schleich2026}

Among quantum algorithms for electronic structure calculations, quantum phase estimation (QPE) provides an asymptotically efficient approach for obtaining eigenvalues with chemical accuracy on fault-tolerant quantum computers.~\cite{aspuru2005simulated} However, its practical success critically depends on the availability of an efficiently preparable trial state with sufficiently large overlap with the target eigenstate. Because runtime scales inversely with this overlap, constructing accurate approximate wavefunctions remains a principal challenge for making QPE practically viable~\cite{fomichev2024initial, berry2025rapid, ollitrault2024enhancing}. Another potential drawback in prospective simulation of static or dynamical properties, where eigenstates must be prepared more than once, is QPE's probabilistic nature. 

Variational Quantum Eigensolvers (VQE)~\cite{peruzzo2014variational, mcclean2016theory} have emerged as a natural framework for obtaining deterministic state-preparation protocols for approximate eigenstates. VQE employs the variational principle by combining a parametrized quantum circuit with a classical optimization routine to minimize the expectation value of the electronic Hamiltonian. The optimization is costly (high shot-count, many individual measurements)~\cite{ghontier2022measurements,patel2025quantum}, but once obtained, state preparation with the optimized circuit is comparably cheap. Besides being a leading algorithm for research near-term quantum hardware in the last decade,~\cite{cerezo2021variational, anand2022quantum, tilly2022Variational} a significant advantage to classical counterparts originating from VQEs alone is unlikely; they can, however, play an important role in the fault-tolerant era by generating high-quality initial states, or preparing near-exact states for systems, or parts of them~\cite{santosHybrid2025, santos2026shallow}, with less correlated ground states. 

Despite the success of variational methods, the accompanying classical optimization remains a major bottleneck. The optimization landscape can exhibit barren plateaus~\cite{mcclean2018barren, Mao2024} and numerous local minima, making reliable convergence increasingly difficult for large or unstructured circuit designs. These challenges have motivated alternative optimization strategies, including machine learning approaches that predict or initialize variational parameters, reducing optimization cost and improving robustness.~\cite{vpm, mioa2024neural, bensoussan2025accelerq, related-warm-starting-vqe}

A wide variety of variational ansätze have been proposed over the years~\cite{grimsley2019adaptive, burton2023Exact, burton2024Accurate, anand2022quantum}. However, many either require deep quantum circuits, involve a large number of variational parameters, or lack the robustness necessary for black-box application across diverse chemical systems. One notable exception is the separable pair approximation (SPA)~\cite{kottmann2022optimized}, which has demonstrated consistently strong performance across a broad range of applications. In particular, SPA has served as a reliable baseline for benchmarking variational algorithms~\cite{weber2022toward,schleich2023partitioning} and has successfully been applied to target diverse molecular properties~\cite{gil2025sharc,santosHybrid2025, santos2026shallow}. The method computes a shallow circuit with constant depth in system size and linear depth w.r.t. the basis functions per electron while being classically tractable with cost comparable to mean-field approaches like Hartree-Fock.~\cite{barta2026consistent}
As a result, the method serves as a toy model for exploring larger systems or as an initial building block for more advanced circuit designs.~\cite{kottmann2024quantum, kottmann2023Molecular}

In prior work~\cite{vpm}, it was shown that graph convolutional neural networks (GNNs) can predict the
optimal parameters of SPA circuits accurately, extrapolating from a small
training set to larger system sizes (number of atoms) and varying structures
(linear chains, rings and amorphous arrangements). These models were
restricted to hydrogenic systems, that is, systems $\mathrm{H}_n$ of $n$
hydrogen atoms. In this work, we ask whether this transferability extends to
general molecules without retraining. 
To this end, we develop an alchemical approach to transfer hydrogenic systems to organic molecules. We demonstrate its success with SPA circuits and show first results that extend to quantum subspace designs using multiple circuits.

\section{\label{methodology}Methodology}

We begin from an assumption rather than from a construction. Suppose we are given a black box $\mathcal{S}$, for any small hydrogenic system $h$, combined with a suitable parametrized quantum circuit $U(\theta)$, returns the optimal parameters of the quantum circuit w.r.t ground state preparation
\begin{equation}
\label{eq:solver}
    \mathcal{S}:\;h\;\longmapsto\;\theta_h^\star ,
\end{equation}
such that
\begin{equation}
    \theta_h^\star = \arg\min_{\theta} \bra{0}U(\theta)^\dagger H(h) U(\theta)\ket{0},
\end{equation}
with $H(h)$ being the Hamiltonian of the hydrogenic system $h$. 
In practice, it is sufficient to have an approximator $\mathcal{S}$ for the optimal circuit angles. 
In a prior work, such an approximator was, for example, constructed for SPA circuits using graph neural networks.~\cite{vpm} 
In this case, the approximator was restricted to hydrogenic systems, as they are conceptually simpler than general molecules and handling them required less ``chemical intuition''.~\cite{barta2026consistent}

In this work, we will focus on the transferability of the predictor $\mathcal{S}$ to non-hydrogenic molecules. We show that in many cases we can efficiently transfer a general molecular instance $\mathcal{M}$ to a suitable hydrogenic instance via a map $\mathcal{P}(\mathcal{M})=h$ and get good approximations for the optimal parameters via
\begin{equation}
    \theta^\star = \mathcal{S}(\mathcal{P}(\mathcal
    M)).
\end{equation}
We will give explicit constructions for the transfer operation $\mathcal{P}$, illustrate examples for $\mathcal{S}$, and analyze shortcomings and limitations.

The cost of the resulting procedure is that of a handful of independent calls to $\mathcal{S}$ on small systems, without necessarily ever performing an optimization on the target Hamiltonian $H(\mathcal{M})$. Two ingredients have to be supplied to make this statement operational, and these constitute the contribution of this section:
\begin{enumerate}
    \item A \emph{projection} $\mathcal{P}:\;\mathcal{M}\rightarrow\{h_1,\dots,h_F\}$ that maps a target molecule $\mathcal{M}$ onto hydrogenic fragment systems $h_f$ and assigns the orbitals of $\mathcal{M}$ to exactly one fragment (\autoref{sec:projection}, \autoref{sec:decomposition}).
    \item A \emph{transfer} rule, which here is simply the concatenation $\tilde{\theta}=\theta_1\oplus\dots\oplus\theta_F$ of the variational parameters returned by $\mathcal{S}$, inserted into the target circuit of $\mathcal{M}$ and evaluated against the full molecular Hamiltonian $H(\mathcal{M})$.
\end{enumerate}
The resulting state $\ket{\Psi(\tilde{\theta})}$ is used either directly as an approximation to the target wavefunction, or as an initial state for a subsequent variational optimization on $H$.

\subsection{\label{sec:projection}Hydrogenic Projection}

\begin{figure*}[t]
    \centering
    \includegraphics[
    % trim=0 2cm 0 2cm,
    clip,
    width=0.95\textwidth]{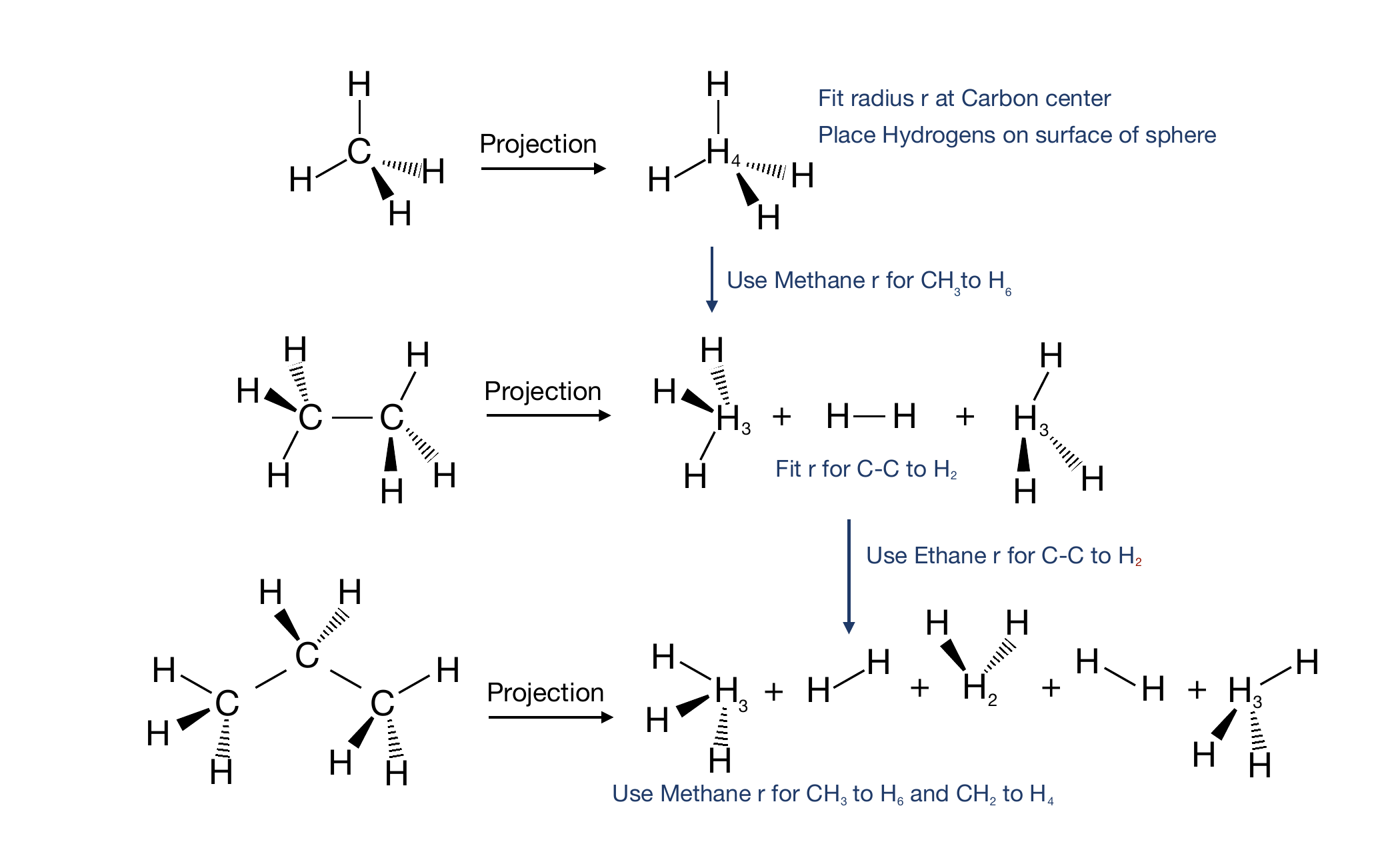}
    \caption{Protocol overview for decomposing the molecules methane, ethane and propane into their hydrogenic projections. Each newly introduced connection type requires a single radius fit, which is subsequently transferred unchanged to any larger molecule containing the same connection type.}
    \label{fig:overview}
\end{figure*}

The projection $\mathcal{P}$ is constructed by determining the hydrogenic connectivity needed to reproduce the target molecule's local bonding environment. Surrogate hydrogen atoms are placed on the surface of a sphere centered on the nucleus under consideration, along the bond directions of the original molecule; the projection radius $r$ is the only free parameter of the map. It is fixed once per bond type by maximizing the overlap between the target wavefunction $\ket{\psi_t}$ and the hydrogenic wavefunction $\ket{\psi'(r)}$,
\begin{equation}
\label{eq:r}
    r^\star=\argmax_r \left|\braket{\psi_t|\psi'(r)}\right|^2.
\end{equation}
Here, $r$ denotes the radius of the projected sphere centered on the atomic nucleus, as illustrated for methane in \autoref{fig:projection_methane}. Because $r$ is attached to a bond type rather than to a molecule, a radius fitted on a small reference system is reusable in every larger molecule containing the same bond -- meaning, we will fit the radius for methane and then leave it as a fixed parameter for carbon atoms in a methane-like ($sp^3$) structure.
This is how the protocol generalizes, as illustrated in \autoref{fig:overview}: a single radius fixes the C--H channel in methane, and the same radius is reused unchanged for ethane and propane. Molecules containing several distinct bond types simply carry several radii, one per connector type of the composing atoms.

The surrogate hydrogens are conceptually related to the capping atoms of
hybrid QM/MM and ONIOM schemes, in which a severed bond is saturated by a
link atom, usually hydrogen, placed along the original bond
direction~\cite{chung2015oniom}, and to pseudobond and quantum capping
potential approaches, in which the boundary atom is replaced by a
one-free-valence atom tuned to reproduce the length and strength of the
original bond~\cite{zhang1999pseudobond,dilabio2002capping}. The projection
radius $r$ plays the role of the link-atom placement factor in these schemes.
The difference lies in the target quantity: the surrogate atoms here are not
used to compute energies of a partitioned system, but to define a hydrogenic
stand-in whose variational circuit parameters are transferred.

This differs in purpose from fragmentation and embedding schemes combined
with VQE, such as DMET-VQE~\cite{kawashima2021dmet} or Deep
VQE~\cite{fujii2022deepvqe}, which use fragments to approximate the
energy of the full system. Here the fragments serve only to supply
parameters, where the energy is always evaluated with the full molecular
Hamiltonian.

\subsection{\label{sec:decomposition}Fragment Decomposition and Parameter Transfer}

We can further exploit the locality of the approximation by decomposing the hydrogenic mapping into weakly interacting subsystems and determining the corresponding variational parameters independently. Instead of constructing a single large hydrogenic model, we solve each chemically meaningful fragment separately, preserving the accuracy of the local approximation and removing the need for global optimization.

Take ethane $C_2H_6$ as an example: it decomposes into two methyl (\(\mathrm{CH}_3\)) fragments joined by a single C--C bond, as seen in \autoref{fig:overview}, giving rise to two hydrogenic \(\mathrm{H}_6\) subsystems and one \(\mathrm{H}_2\) subsystem, each of which is solved independently. The resulting parameters provide a more accurate initialization for the complete molecular wavefunction than a global fit would, since the approximation error remains confined to the weak interactions between fragments rather than accumulating over the entire system. Moreover, because the projection radius of each subsystem is optimized while accounting for the interactions among all atoms within that fragment, local approximation errors partially compensate for one another.

Note that the protocol is useful in scenarios where the per-fragment errors do not compound. In \autoref{table:overview}, we provide some numerical demonstrations by computing the transfer error either for an individual minimization of the hydrogen-surrogate (the map $\mathcal{S}$ is a minimizer) or for predicted angles (the map $\mathcal{S}$ is the predictor constructed in the following section).

The complete procedure is summarized as follows:
\begin{enumerate}
    \item \textbf{Fit.} Determine the projection radius of every interatomic connection type from a small reference molecule via \autoref{eq:r}.
    \item \textbf{Decompose.} Split the target molecule into the hydrogenic subsystems induced by these projections.
    \item \textbf{Solve.} Invoke the black box $\mathcal{S}$ on each decoupled subsystem.
    \item \textbf{Transfer.} Insert the collected parameters into the target circuit of the target molecule and evaluate against its original Hamiltonian.
\end{enumerate}
Only step 3 depends on the choice of $\mathcal{S}$, and only through the accuracy and cost of that choice.

\subsection{\label{sec:ansatz}Circuit Designs}
 
The protocol developed in this section provides variational parameters that depend on the chosen circuit.
Here we will use two variational algorithms: a single-circuit classically tractable approach (SPA) and a multi-circuit subspace approach (MCBV).
\begin{enumerate}
    \item SPA: shallow and classically simulable~\cite{barta2026consistent}, suitable approximators $\mathcal{S}$ exist~\cite{vpm}. A computationally cheap testbed for the proposed hydrogenic projections.
    \item MCVB: more expressive than SPA but still based on chemical principles~\cite{kottmann2024quantum}, likely not to be classically simulable; parameter optimization is the bottleneck in quantum computational approaches.
\end{enumerate}
In summary, SPA circuits are candidates whose structure naturally suits the fragmentation approach described in the previous section; they are comparably cheap to simulate and will serve as the primary testbed for the transfer strategies presented here. In contrast, MCVB circuits represent a use case where transfer learning can have a significant impact.
Next, we briefly recap both methods, focusing only on the essential parts for the context of this work. 

\subsubsection{\label{sec:spa_ansatz}Separable Pair Ansatz}

The separable pair approximation (SPA) writes the many-electron wavefunction
as a product of $N_e/2$ independent electron-pair functions $\ket{\psi_k}$,
each living in its own disjoint set of orbitals,
\begin{equation}
\label{eq:spa_state}
    \ket{\Psi_{\mathrm{SPA}}} = \bigotimes_{k=1}^{N_e/2} \ket{\psi_k} .
\end{equation}
In this work, each pair is shared between two orbitals. This assignment is
naturally expressed as a resonance graph whose vertices are the orbitals and whose
edges $e=(i,j)$ connect the two orbitals of each pair. Since the orbitals are
localized, every edge is associated with the two atomic centers it connects,
so that each edge corresponds to a chemical bond and the full graph to one
valence-bond structure of the molecule. We therefore label each pair function
and its angle by its edge, $\ket{\psi_e}$ and $\theta_e$.
The circuit design is therefore especially suitable for the fragmentation approach described above.

On a quantum computer, each pair function is prepared by a short circuit of
one- and two-electron excitation gates controlled by its single angle
$\theta_e$. The parameter count therefore grows only linearly with the number
of electron pairs, and the circuit stays shallow. The angles are found by
minimizing the energy expectation value in a standard VQE loop. Beyond being a
compact ansatz on its own, SPA serves as a building block for more expressive
circuits, such as the multiconfigurational valence-bond ansatz in the next
section.
 
\subsubsection{\label{sec:mvb}Multiconfigurational Valence-Bond Ansatz}
 
The multiconfigurational valence-bond (MCVB) ansatz constructs the molecular
wavefunction as a linear combination of $N$ parametrized circuits,
\begin{equation}
\label{eq:mcvb_state}
    \ket{\Psi(\mathbf{c},\boldsymbol{\theta})}
    = \sum_{k=1}^{N} c_k\, U_k(\theta_k)\ket{0} ,
\end{equation}
where each circuit prepares an individual valence-bond configuration
$\ket{\psi(\theta_k)} = U_k(\theta_k)\ket{0}$ expressed by a resonance graph (see above). 
Taking the $U_k$ to be SPA circuits over different molecular
graphs recovers the chemical picture of resonance: each term is one Lewis
structure, and their superposition describes the delocalization that no single
structure can. The choice of the circuit family $U_k$ therefore determines the
expressiveness of the ansatz and, consequently, the achievable accuracy.
 
The expansion coefficients $\mathbf{c}$ and the circuit parameters
$\boldsymbol{\theta}$ are optimized simultaneously by minimizing the variational
energy,
\begin{equation}
\label{eq:mcvb_opt}
    (\mathbf{c}^{*},\boldsymbol{\theta}^{*})
    = \argmin_{\mathbf{c},\boldsymbol{\theta}}
    \frac{\bra{\Psi(\mathbf{c},\boldsymbol{\theta})} H \ket{\Psi(\mathbf{c},\boldsymbol{\theta})}}
         {\braket{\Psi(\mathbf{c},\boldsymbol{\theta})|\Psi(\mathbf{c},\boldsymbol{\theta})}} ,
\end{equation}
after which the optimized states span an effective variational subspace in which
the final coefficients follow from the corresponding generalized eigenvalue
problem.

Evaluating the matrix elements poses a classical bottleneck, as no efficient algorithms currently exist.
The optimization procedure for the parameters is, however, a general bottleneck that quantum computation does not lift. On the contrary, on quantum hardware this is not expected to become better, as gradient evaluation (even with noiseless execution) will become even more expensive due to finite sampling errors. Predicting the parameters directly is a possible route to feasible algorithms.

\subsection{\label{init}The Hydrogenic Subproblem}

The protocol of \autoref{methodology} is complete except for the black box $\mathcal{S}$ of \autoref{eq:solver}. This section introduces a cheaper model for SPA circuits complementary to the existing GNN approximator of Ref.~\cite{vpm}. It is self-contained: nothing in \autoref{sec:projection} or \autoref{sec:decomposition} depends on it, and it may be replaced by any of the alternatives mentioned above, without altering the protocol.

We employ a one-shot, purely distance-based model that predicts each SPA-circuit
parameter $\theta_e$ directly from its edge $e$
(Sec.~\ref{sec:spa_ansatz}), specifically from its edge weight $x(e)$ that we chose to be the distance between the two centers
it connects, without any optimization at inference time. It is chosen here because, in contrast to the graph neural networks, it is conceptually simple and computationally cheap to train: each parameter follows from a single closed-form function of one interatomic distance.
Concretely, for an edge $E_i = (u,v)$ of the molecular graph $G$, we model the associated variational parameter as a function of the edge length, i.e.\ the Euclidean distance between the two incident atoms,
\begin{equation}
\label{eq:edgelength}
x(\left\{u,v\right\}) = \|\mathbf{x}_u - \mathbf{x}_v\|_2 ,
\end{equation}
where $\mathbf{x}_u,\mathbf{x}_v \in \mathbb{R}^3$ denote the real-space coordinates of the atoms $u$ and $v$ -- we can view the molecular graph as a weighted graph $(V,G,x)$ with atoms $V$ edges $E$ and edge weights $x$. 
In the following we will develop a simple model that transfers these edge weights $x$ into circuit parameters $\sigma: x\rightarrow \theta$.
We incorporate boundary conditions into the model $\sigma$: it approaches $0$ at short interatomic distances and $-\pi/2$ at large separations, where the pair dissociates. The latter reflects the true behaviour at dissociation, the former the model limit, as we will exclude edge cases $x_u = x_v$ with infinite coulombic repulsion of the atoms. To interpolate smoothly between these limits we use the standard sigmoid
\[
\sigma(x) = \frac{1}{1 + e^{-x}}
\]
as a fundamental building block, and parametrize the SPA parameter of the edge as
\begin{equation}
\label{eq:sig1}
    \sigma_{\vec{p}}(x) = -\frac{\pi}{2}\,\sigma(p_1 - p_2\,x),
\end{equation}
which reproduces both limits by construction and is purely a function of the single distance $x$ for a given edge. The two free parameters $\vec{p} = (p_1, p_2)$ set the location and sharpness of the crossover.

The parameters $\vec{p}$ are obtained \textit{once} by regression on the hydrogen dimer $\mathrm{H}_2$: we sample $100$ interatomic distances between $0.5\,\text{\AA}$ and $5.0\,\text{\AA}$, compute the corresponding optimized SPA parameters, and fit \autoref{eq:sig1} to this data until convergence.
The fitted saturation curve is reused for every
edge, so that the solver $\mathcal{S}$ reduces to a single evaluation of
\autoref{eq:sig1} per bond. Note, that this is rather a pragmatic than a rigorous choice as the dependence could in principle be solved analytically. The angles are therefore obtained in one shot: no
iteration, no fragment diagonalization, and no Hamiltonian access beyond the
interatomic distances themselves, at a low computational cost for the entire system.\\

In comparisson to Meta-VQEs, which learns the parameters as a
function of a Hamiltonian parameter from a few training
points~\cite{cervera2021metavqe}, the sigmoid of \autoref{eq:sig1} can be read
as a minimal, two-parameter instance of this idea, but restricted to a single
interatomic distance and trained once on $\mathrm{H}_2$ geometries.

Within the class of SPA circuits limited to hydrogenic systems this fullfills all required properties for an approximator $\mathcal{S}$.
Beyond, we can use it as an empirical tool to initialize SPA parameters for a full optimization, which gives us a fairly cheap way to obtain formally exact values and simulate a perfect predictor $\mathcal{S}$.
In the appendix, \autoref{table:diag} quantifies both regimes on hydrogenic systems: the
one-shot error, i.e.\ the accuracy of the predicted angles before any
optimization is performed, and the number of iterations required to reach
convergence when the optimizer is started from them.

\section{\label{sec:exp}Experiments}

WE used generic hydrocarbons as a testbed for evaluating our approach. Since each carbon atom forms four connectors, the projection
radii associated with C--H and C--C bonds can be determined from methane and
ethane, respectively, and subsequently transferred to larger molecules without
refitting. This establishes a bottom-up protocol in which the molecular size is
increased gradually while the parameter set remains fixed, as outlined in
\autoref{fig:overview}.\newline

\paragraph*{Computational details.}
For SPA circuits, all calculations are carried out in the minimal STO-3G basis, using the
hardcore-boson representation of the electronic Hamiltonian throughout. 
In hydrogenic systems, the orbitals are heuristically initialized to reflect the $H-H$ bonds (see section III.1 of ~\cite{barta2026consistent}) and similarly for the hydrocarbons (see section III.2 of ~\cite{barta2026consistent}) where we employed "Chemist’s
localized property-optimized orbitals" (CLPOs).\cite{nikolaienko2019Localizeda, nikolaienko2014JANPA}. We refer to Ref. ~\cite{vanderhorst2026transferablemachinelearningapproach} for initial approaches to predict the optimal orbital coefficients, and Ref.~\cite{langkabel2024advent} on how to go beyond STO-3G accuracy. 
All circuits are simulated classically with the \texttt{tequila} framework~\cite{kottmann2021tequila}, and the
baseline SPA optimizations use a two-point gradient estimator, since in the current implementation analytical
gradients became a bottleneck for the larger instances. Unless stated
otherwise, reported errors are taken with respect to a fully converged SPA
optimization performed directly on the target Hamiltonian, so that the quoted
values quantify the transfer error of the protocol and not the intrinsic
error of the SPA ansatz; the latter is assessed separately against FCI in
\autoref{Propane} to give the reader some guidance on the quality of the circuit design. As a reference scale, we use
$1\,\mathrm{kcal/mol}=1.594\,\mathrm{mE}_h$, which is the infamous \textit{chemical accuracy}, the required energy precision for {relative} energies in order to predict kinetics accurately, which does not directly translate to our setting (absolute energies in model chemistries) but can still be viewed as a good guideline for accuracy. For simplicity and consistency with the literature, we will stick to atomic units throughout.

\subsection{\label{MethaneEthane}Training Systems: Methane and Ethane}

\begin{figure}[t]
    \centering
    \includegraphics[
    trim=0 0cm 0 1.15cm,
    clip,
    width=0.8\linewidth
]{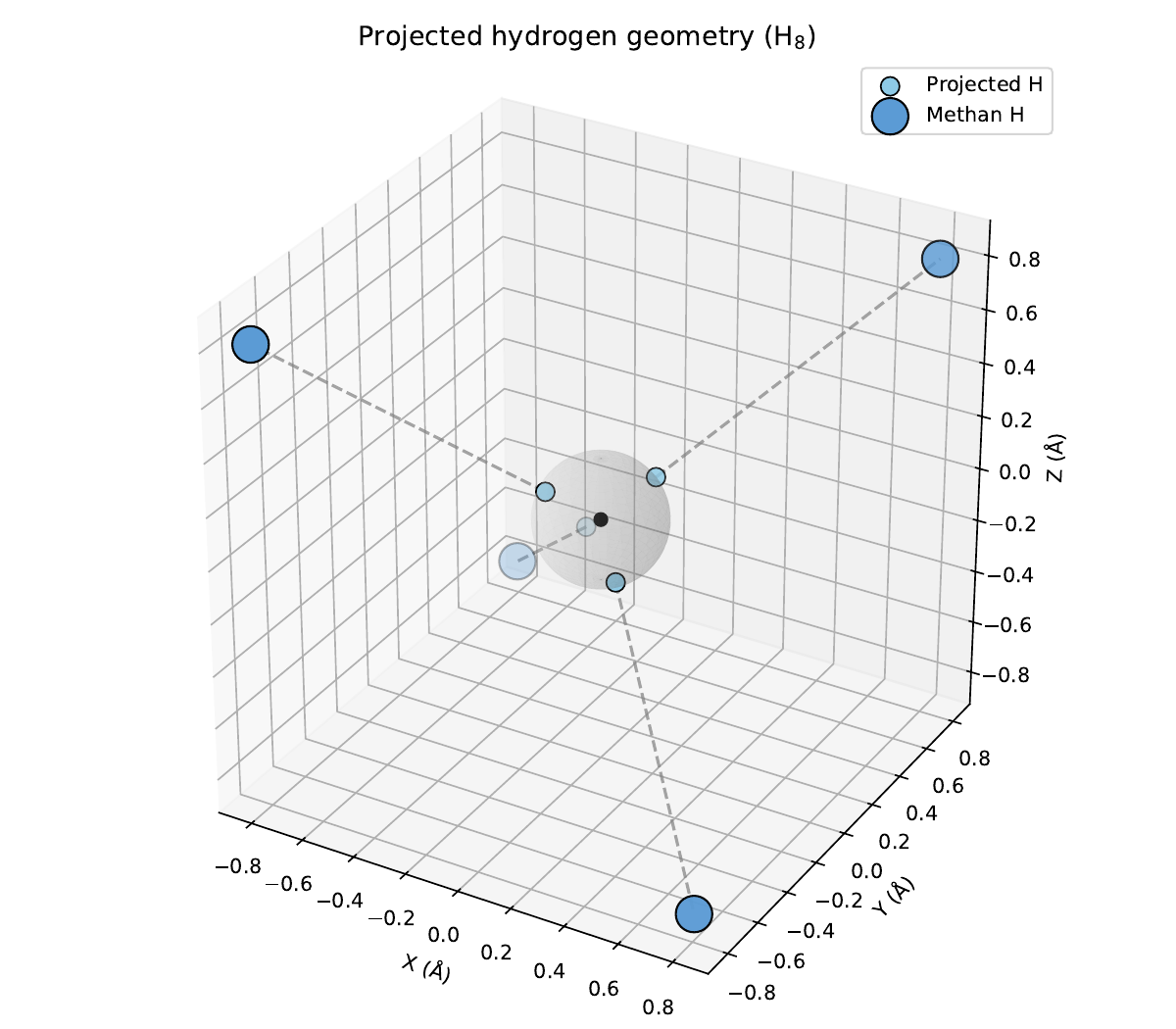}
    \caption{Projection of surrogate hydrogen atoms onto a sphere of radius $r$
    centered at the carbon position (black dot), mimicking methane by an
    $\mathrm{H}_8$ system.}
    \label{fig:projection_methane}
\end{figure}

For methane, the C--H connections are approximated by projecting surrogate
hydrogen atoms onto a sphere centered at the carbon position, as illustrated in
\autoref{fig:projection_methane}. This replaces the molecule by a hydrogenic
modelling problem with fixed connectivity: each projected hydrogen is bonded to
its counterpart, so that the molecular graph is preserved.

The difference in interaction behaviour between C--H and H--H bonds is absorbed
into the single effective parameter $r$ of \autoref{eq:r}, obtained by
maximizing the overlap between the hydrogenic and the methane wavefunction. We
emphasize that this is a one-parameter fit per bond type, performed once on the
reference molecule; no further fitting is carried out for any larger system.

\begin{figure}[t]
    \centering
    \includegraphics[width=0.85\linewidth]{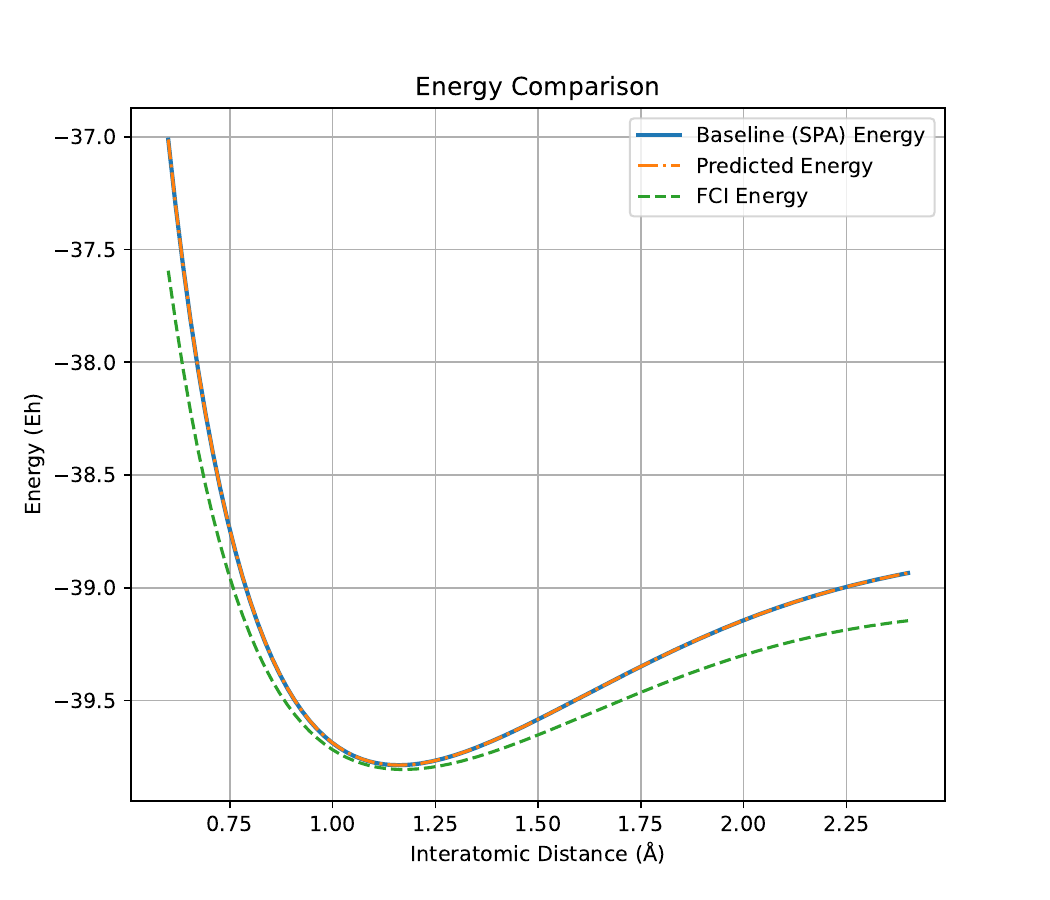}
    \caption{Methane Training Evaluation: Energy comparison between the projected molecular energies and the
    baseline SPA optimization results, achieving a mean error of
    $0.02\,\mathrm{mE}_h$.}
    \label{fig:methane_energy}
\end{figure}

We generate 100 molecular configurations by dissociating the outer hydrogen
atoms radially outward at constant angular separation. For each configuration we
minimize the circuit parameters on the projected $\mathrm{H}_8$ geometry and
evaluate the resulting wavefunction with the methane Hamiltonian directly,
without any further optimization. The resulting energies are compared against
conventional SPA optimization baselines in \autoref{fig:methane_energy}. Across
the 100 instances the mean transfer error is $0.02\,\mathrm{mE}_h$, which is more than acceptable with repsect to the target accuracy of one millihartree.

The transition from methane to ethane introduces the carbon--carbon interaction
as an additional degree of freedom. To capture it, we introduce a second
spherical projection centered at the carbon atoms, which parametrizes the C--C
coupling independently of the previously fitted C--H interaction. The two
channels are treated on equal footing but carry distinct radii, each obtained
from a separate one-parameter fit via \autoref{eq:r}. Ethane is the last system
for which any fitting is performed.

To quantify the performance of the model, we evaluate the contribution of each
interaction channel to the total energy prediction error separately: once with
only the C--H bonds projected, once with only the C--C bond projected, and once
with both channels projected simultaneously. This decomposition is reported for
all molecules in \autoref{table:overview}; for ethane it yields
$0.05\,\mathrm{mE}_h$ and $0.07\,\mathrm{mE}_h$ for the individual channels
and $0.13\,\mathrm{mE}_h$ when both are applied. Again, this is more than acceptable.

\subsection{\label{Propane}Evaluation Systems: Propane and Norbornane}

\begin{figure}[t]
    \centering
    \includegraphics[width=0.85\linewidth]{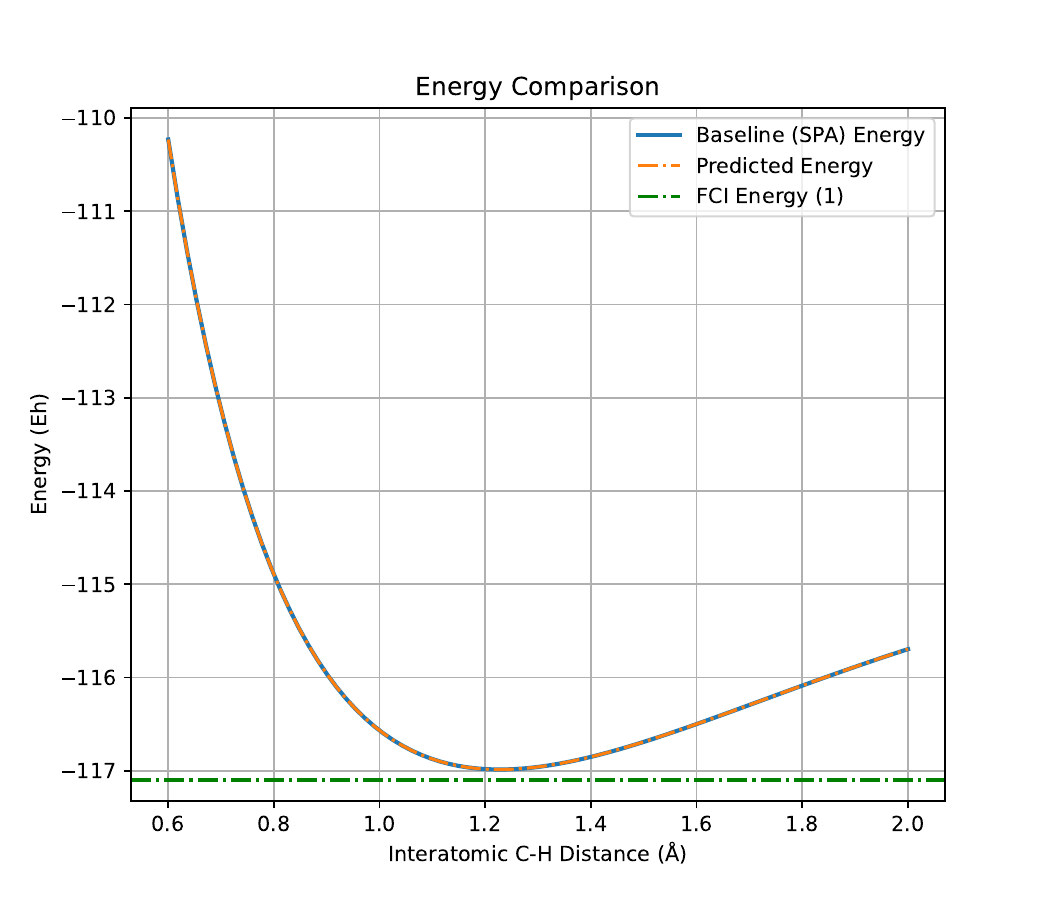}
    \caption{Energy comparison between the projected molecular energies and the
    optimization baselines for propane. The model achieves a mean error of
    $0.181\,\mathrm{mE}_h$. The full configuration interaction (FCI) energy is
    shown only for the molecular configuration with lowest energy.}
    \label{fig:propane_energy}
\end{figure}

Propane is the first system evaluated entirely without introducing additional
fitted parameters and therefore constitutes a direct test of the generalization
capability of the projection framework (\autoref{fig:propane_energy}). The mean
transfer error over the generated configurations is $0.18\,\mathrm{mE}_h$.

It is instructive to separate this quantity from the intrinsic error of the SPA
ansatz itself. For the equilibrium-like configuration, the deviation of the
projected energy from the FCI reference decomposes as
\begin{equation}
\label{eq:error_split}
\underbrace{E_{\mathrm{proj}} - E_{\mathrm{FCI}}}_{113.28\,\mathrm{mE}_h}
=
\underbrace{E_{\mathrm{proj}} - E_{\mathrm{SPA}}^{*}}_{\text{transfer error}}
+
\underbrace{E_{\mathrm{SPA}}^{*} - E_{\mathrm{FCI}}}_{\text{ansatz error}} ,
\end{equation}
where $E_{\mathrm{SPA}}^{*}$ denotes the fully converged SPA energy and the
total amounts to around $113\,\mathrm{mE}_h$ in the STO-3G basis.

Given the magnitudes reported in \autoref{table:overview}, the residual error is
dominated by the limited expressiveness of the SPA ansatz rather than by the
parameter transfer. Improving upon this figure therefore requires a more
expressive ansatz rather than a more accurate projection. We will follow this direction
in \autoref{sec:mvb}.

Computing the exact FCI energy for propane is computationally
demanding due to the exponential growth of the Hilbert space, involving
approximately $1.2\times10^{12}$ determinants; the reference value was taken
from Ref.~\cite{fci}. By contrast, the proposed method obtains its energy from
several decoupled hydrogenic subsystems, without iterative optimization on the
target Hamiltonian and without diagonalization in the full determinant space.

To conclude the evaluation on hydrocarbons we apply the protocol to norbornane,
which, unlike the previously considered systems, possesses a rigid bicyclic
framework of seven carbon atoms. Its topology provides a stringent test of the
transferability of the fitted radii, since the C--C connectors now occur in
strained ring environments absent from the acyclic reference molecules.

The transfer error remains below the millihartree target accuracy at $0.55\,\mathrm{mE}_h$,
but the C--C channel degrades noticeably: its contribution grows from
$0.07\,\mathrm{mE}_h$ in ethane and $0.11\,\mathrm{mE}_h$ in propane to
$0.43\,\mathrm{mE}_h$ in norbornane, whereas the C--H channel grows only
mildly. We attribute this to the ring strain, which places the C--C bonds
outside the geometric regime in which the radius was fitted, and it indicates
that strained topologies are the first place where a refit or a more
flexible projection could become necessary.

\begin{table}[t]
\centering
\small
\setlength{\tabcolsep}{5pt}
\begin{tabular}{llcc}
\toprule
\textbf{Molecule} & \textbf{Prediction} & \textbf{E$_{\text{min}}$ (mE$_h$)} & \textbf{E$_{\text{pred}}$ (mE$_h$)}\\
\midrule 
\multicolumn{3}{c}{\vspace{-0.9em}} \\[0.15em]
Methane & C-H &  0.022 & 0.124\\
\multicolumn{3}{c}{\vspace{-0.9em}} \\[0.15em]
\cmidrule(lr){1-4}

\multirow{3}{*}{Ethane}
 & C-H  & 0.049 & 0.364\\
 & C-C    & 0.074 & 0.866 \\
 & Both  &  0.131 & 1.230\\
\cmidrule(lr){1-4}

\multirow{3}{*}{Propane}
 & C-H  & 0.072 & 0.387\\
 & C-C    & 0.108 & 1.785\\
 & Both  & 0.181 & 2.171\\
\cmidrule(lr){1-4}

\multirow{3}{*}{Norbornane}
 & C-H  & 0.128 & 1.983\\
 & C-C    & 0.425 & 4.320 \\
 & Both  & 0.553 & 6.301 \\
\bottomrule
\end{tabular}
\caption{Transfer error of the projected wavefunction evaluated with the
original molecular Hamiltonian, measured against a fully converged SPA
optimization of the same molecule. Each channel row is an independent
calculation in which only the indicated bond type is projected.}
\label{table:overview}
\end{table}

In \autoref{table:overview} we also compare solving the hydrogenic mapping via minimization with the model as initial guess (E$_{\text{min}}$) against using the regression model directly (E$_{\text{pred}}$) on the target molecules. Here we want to stress that this prediction process only requires the two parameters for the H$_2$ regression and one parameter for the bond type, resulting in a very low parameter model.

\subsection{\label{sec:convergence}Convergence Behaviour}

Using the projected wavefunctions summarized in \autoref{table:overview} as
initial states (starting from predicted angles of the model introduced earlier) for the variational optimization, we evaluate their impact on the
convergence of the SPA optimization. Specifically, we compare the average number
of circuit evaluations required by the optimizer to reach convergence when
initialized from the default parameter values $\theta = 0$ and when initialized with the
parameters obtained from the proposed projection protocol. The results are
summarized in \autoref{table:iterations}.

\begin{table}[t]
\centering
\small
\setlength{\tabcolsep}{8pt}
\begin{tabular}{lcc}
\toprule
\textbf{Molecule} & \textbf{Default} & \textbf{Projection} \\
\midrule
Methane    & 18  & 12 \\
Ethane     & 39  & 22 \\
Propane    & 69  & 30 \\
Norbornane & 103 & 56 \\
\bottomrule
\end{tabular}
\caption{Average number of circuit evaluations required for the SPA
optimization to converge using the default initialization and the proposed
projection-based initialization. We used the comparable cheap prediction model introduced in this work for the initialization protocol. }
\label{table:iterations}
\end{table}

The values reported for the projection-based initialization exclude the circuit
evaluations spent on the decomposed hydrogenic subsystems. These are not
negligible in number, but they are incurred on drastically smaller problems, so
that counting circuit evaluations alone weights a fragment evaluation and a
full-molecule evaluation equally and therefore understates the advantage rather
than overstating it. A faithful comparison requires a cost metric that accounts
for the system size of each optimization problem.

\providecolor{eA}{RGB}{214,39,40}     % nearest-neighbour bond
\providecolor{eC}{RGB}{31,119,180}    % long-range pairing
\providecommand{\Ng}{N_{\mathrm{g}}}

\subsection{\label{sec:mcvb_results}Beyond Pair Separation: Multiconfigurational Valence-Bond Ansätze}

The protocol developed so far transfers parameters into a single separable state. Its
accuracy is therefore capped by the expressiveness of that ansatz, and
\autoref{Propane} showed that for the alkanes the residual error is dominated by
this cap rather than by the transfer itself. A natural way to lift it is to move
to the multiconfigurational valence-bond (MCVB) ansatz of \autoref{sec:mvb},
which superposes several SPA states built on different molecular graphs. In this
section we apply the protocol to two conjugated $\pi$ systems of increasing
size, butadiene ($\mathrm{C}_4\mathrm{H}_6$) and hexatriene
($\mathrm{C}_6\mathrm{H}_8$).

We stress at the outset that both molecules are treated with the same
learned projection. The projection parameters are those fitted once on the
hydrogenic reference data in \autoref{sec:projection} and \autoref{init}; nothing
is refitted when passing from butadiene to hexatriene, and nothing is refitted
when passing from a single graph to a superposition of several. Hexatriene is
consequently not a second calibration but a pure transfer test, carried out
simultaneously at a larger system size and thus more graphs. 
We refer to Ref.~\cite{santos2026shallow} for strategies to treat these molecules outside the active-space $\pi$-systems. In the following, we stick to the active space and ignore the $\sigma$-framework, since SPA circuits can represent it with decent accuracy and the transfer from hydrogenic systems will work analogously to the previous section.

Restricting the active space to the carbon $2p_z$ orbitals reduces each backbone
to a chain of one-electron centres, so that butadiene maps onto an effective
hydrogenic $\mathrm{H}_4$ system and hexatriene onto $\mathrm{H}_6$. Each
molecular graph $G_g$ is a perfect matching of these centres: every orbital is
paired exactly once, and every edge $e\in G_g$ carries one variational angle
$\theta_e$ supplied by the projection. A graph therefore defines one SPA state
\begin{equation}
\label{eq:vbstate}
    \ket{\Psi_g(\boldsymbol{\theta}_g)}
    = \prod_{e\,\in\,G_g} U_e(\theta_e)\; U_{\mathrm{ref}}\ket{0\cdots0},
\end{equation}
and the MCVB wavefunction is the linear combination of the first $\Ng$ such
states,
\begin{equation}
\label{eq:mcvb_sum}
    \ket{\Psi_{\mathrm{MCVB}}(\Ng)}
    = \sum_{g=1}^{\Ng} c_{g}\,UR_g(\pi/2)\ket{\Psi_g(\boldsymbol{\theta}_g)} ,
\end{equation}
with the coefficients $c_g$ obtained from the generalized eigenvalue problem in
the many-body basis $\{\ket{\Psi_g}\}$. The angles $\boldsymbol{\theta}_g$
enter only through the projection, so increasing $\Ng$ adds structures
without adding any fitted quantity.
The same holds for the orbitals, in all
following examples. Each SPA state is preceded by a fixed orbital rotation
$\mathrm{UR}_g(\pi/2)$ on every edge $e=(i,j)$, which turns the two orbitals of the edge into their bonding and antibonding combination. This rotation is
identical for all graphs and edges and therefore introduces no additional
parameters either. This choice of fixing the orbitals limits expressibility compared to a ``breathing orbitals'' approach (see, for example, Fig.~3 of Ref.~\cite{kottmann2024quantum}) but offers a good trade-off in computational cost for the purpose of this work.
In this work, we will estimate the potential accuracy gain through breathing orbitals by comparison with fully orbital-optimized SPA energies, which are computationally affordable.
\newline

\begin{figure}[H]
    \centering
    \includegraphics[width=0.75\linewidth]{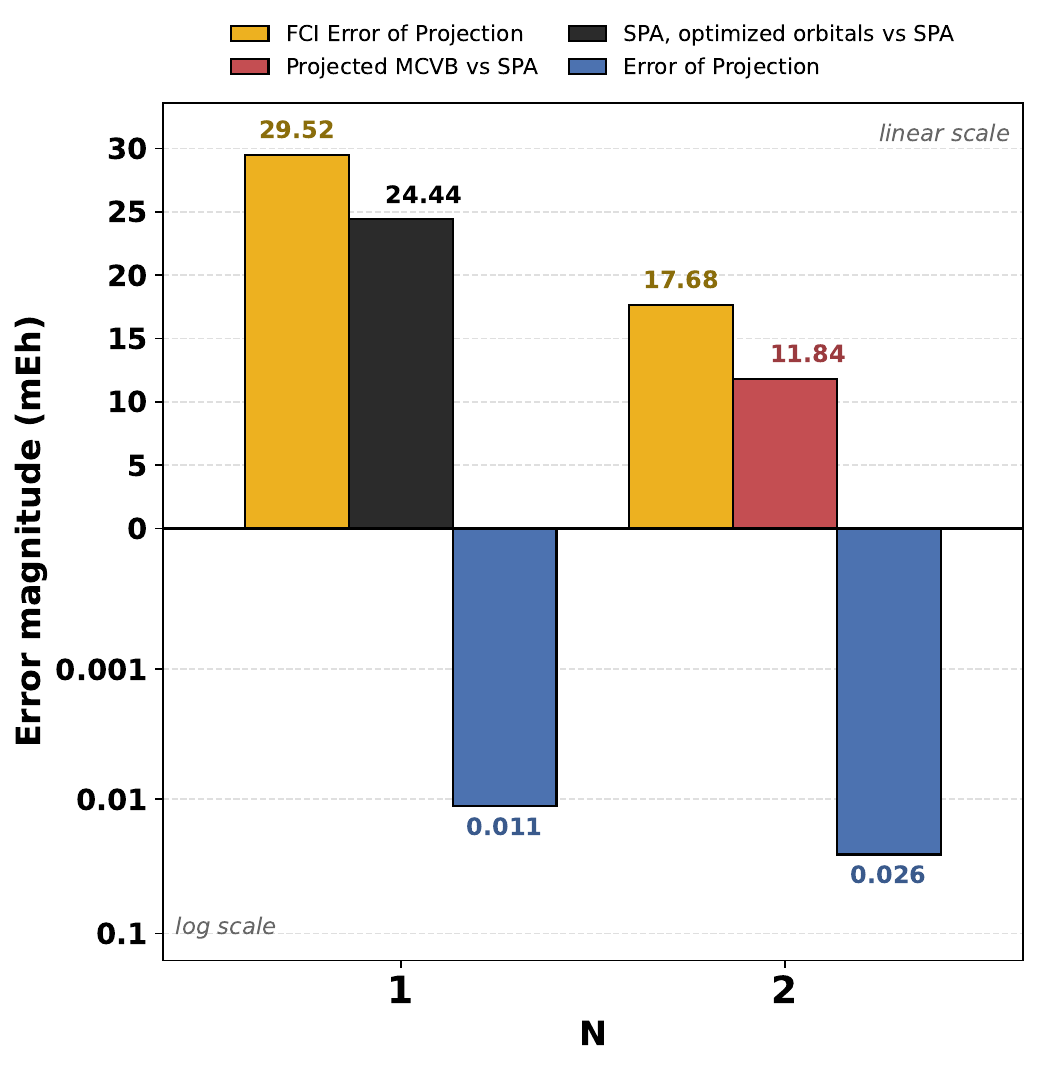}
    \caption{Energy comparison of the minimized MCVB wavefunction with $N$
    molecular graphs using the hydrogenic projection, against the MCVB target
    molecule wavefunction (blue), against the SPA wavefunction (red) and FCI (yellow) .
    SPA with optimized orbitals (black) illustrates the possible gain by using the MCVB Ansatz with free orbital rotations instead of fixed ones. }
    \label{fig:butadine_error}
\end{figure}

\paragraph*{Butadiene.}
For the four-centre $\pi$ system two structures are available, and \autoref{eq:mcvb_sum}
reduces to
\[
\ket{\Psi_{\mathrm{MCVB}}}
=
c_1\,\ket{\butadiene}
+
c_2\,\ket{\butadieneSecond} .
\]
We compare the MCVB energies obtained from the original carbon-chain
representation with those obtained from the projected hydrogenic
$\mathrm{H}_4$ model, evaluating both wavefunctions with the full butadiene
Hamiltonian; the difference between the MCVB and SPA energies then measures the
gain from the second structure. The results are shown in
\autoref{fig:butadine_error}.

The projected MCVB wavefunction lies
below the optimized SPA energy. This is expected rather than anomalous,
since MCVB is strictly more expressive than SPA and can therefore reach lower energies
on the same Hamiltonian. 
Transferring parameters into a higher-level ansatz inherits the accuracy of that
ansatz while retaining the low cost of the hydrogenic projection. The residual
error against the MCVB reference (blue) remains small, confirming that the
transfer itself stays accurate for this circuit.
\newline

\paragraph*{Hexatriene.}
The $\pi$ system of hexatriene is a compact but non-trivial
extension: six electrons in the six carbon $2p_z$ orbitals, on top of a frozen
$\sigma$ core. The six vertices lead to $15$ perfect matchings, but only
$5$ of them (see \autoref{fig:hexgraphs}) are ``allowed'' (non-crossing when the vertices are arranged on a circle) Rumer structures.~\cite{wu2011classicalvbt}

\begin{figure}[h]
    \centering
    \includegraphics[width=0.99\linewidth]{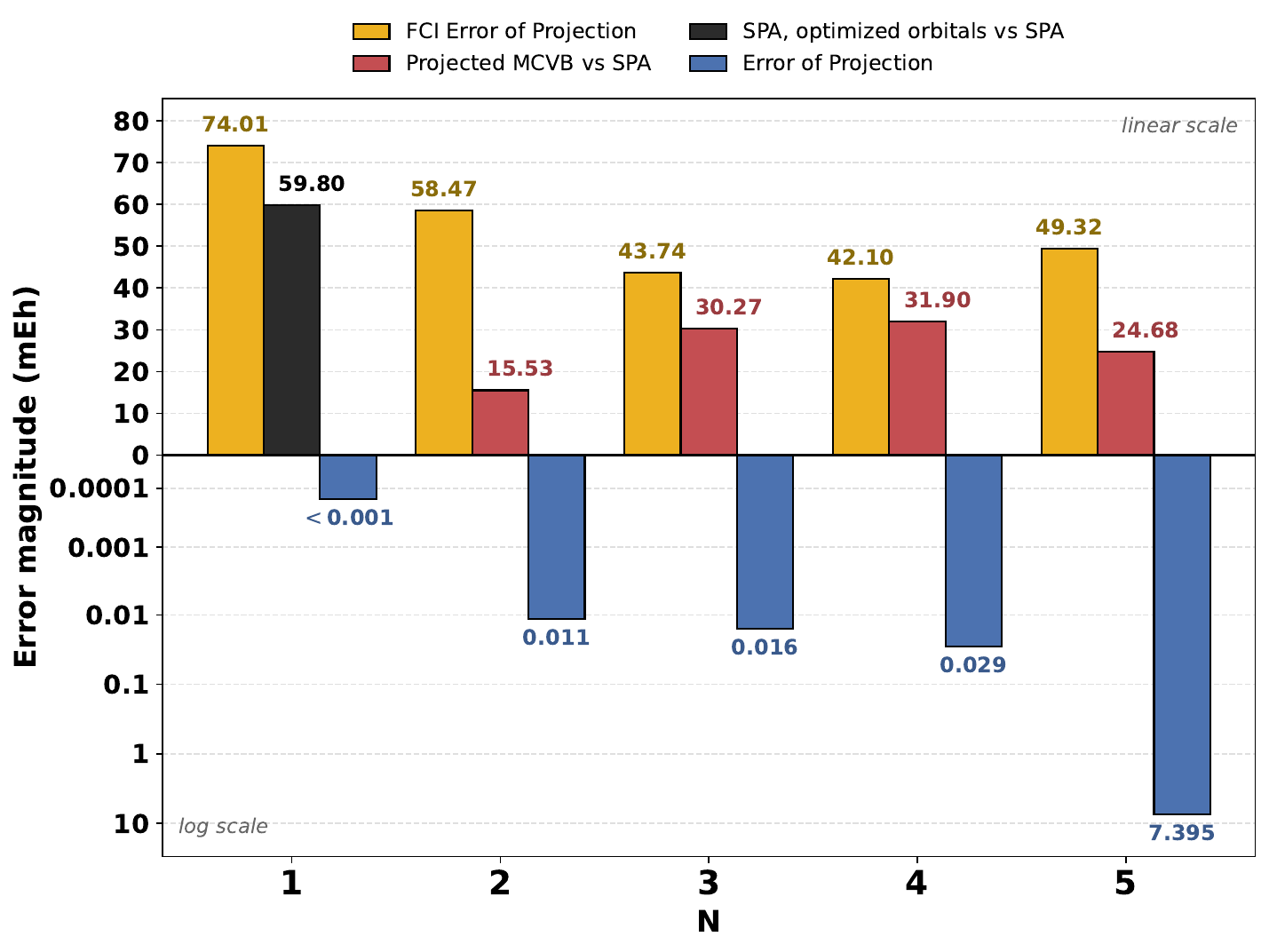}
    \caption{Accuracy of the parameterized MCVB ansatz as a function of the number $N$ of valence bond structures(Fig.~\ref{fig:hexgraphs}). All quantities refer to the projected calculation, in which the parameters are
    transferred to the hydrogenic system. Yellow is the remaining error against
    the FCI energy, red is the gain over the SPA reference, blue is the cost of the transfer
    itself. SPA with optimized orbitals (black) illustrates the possible gain by conjugating the MCVB Ansatz with optimized orbitals. }

    \label{fig:hexresults}
\end{figure}

\autoref{fig:hexresults} collects the errors as structures are added. 
The transfer to the hydrogenic reference system behaves in two clearly separated
regimes. Up to $\Ng=4$ the projection contributes
$0.01$, $0.02$ and $0.03\,\mathrm{mE}_h$, below $0.07\,\%$ of the total
error, so that the projected and directly optimized results are
indistinguishable on the scale of the method error. At $\Ng=5$, however, the
projection error grows by a factor of three orders of magnitude to $7.40\,\mathrm{mE}_h$ and then
accounts for $15\,\%$ of the remaining error. The consequence is visible in
\autoref{fig:hexresults}: while the underlying method keeps improving, reaching
$41.9\,\mathrm{mE}_h$ on the original molecule, the projected energy
deteriorates from $42.1$ to $49.3\,\mathrm{mE}_h$, so that $\Ng=4$ gives the
best projected result.

The structure responsible is $G_5$, the only one carrying two long-range
pairings and hence the most delocalized member of the basis. Its matrix elements
depend most sensitively on the interatomic distances and are therefore the least
transferable, which is consistent with the locality assumption underlying the
projection in \autoref{sec:decomposition}. Transferability thus holds as long as
the added structures remain comparatively local, and degrades for the most
delocalized ones, which is a limitation of the same nature as, the one encountered for LiH in \autoref{limit}.

% ======================================================================
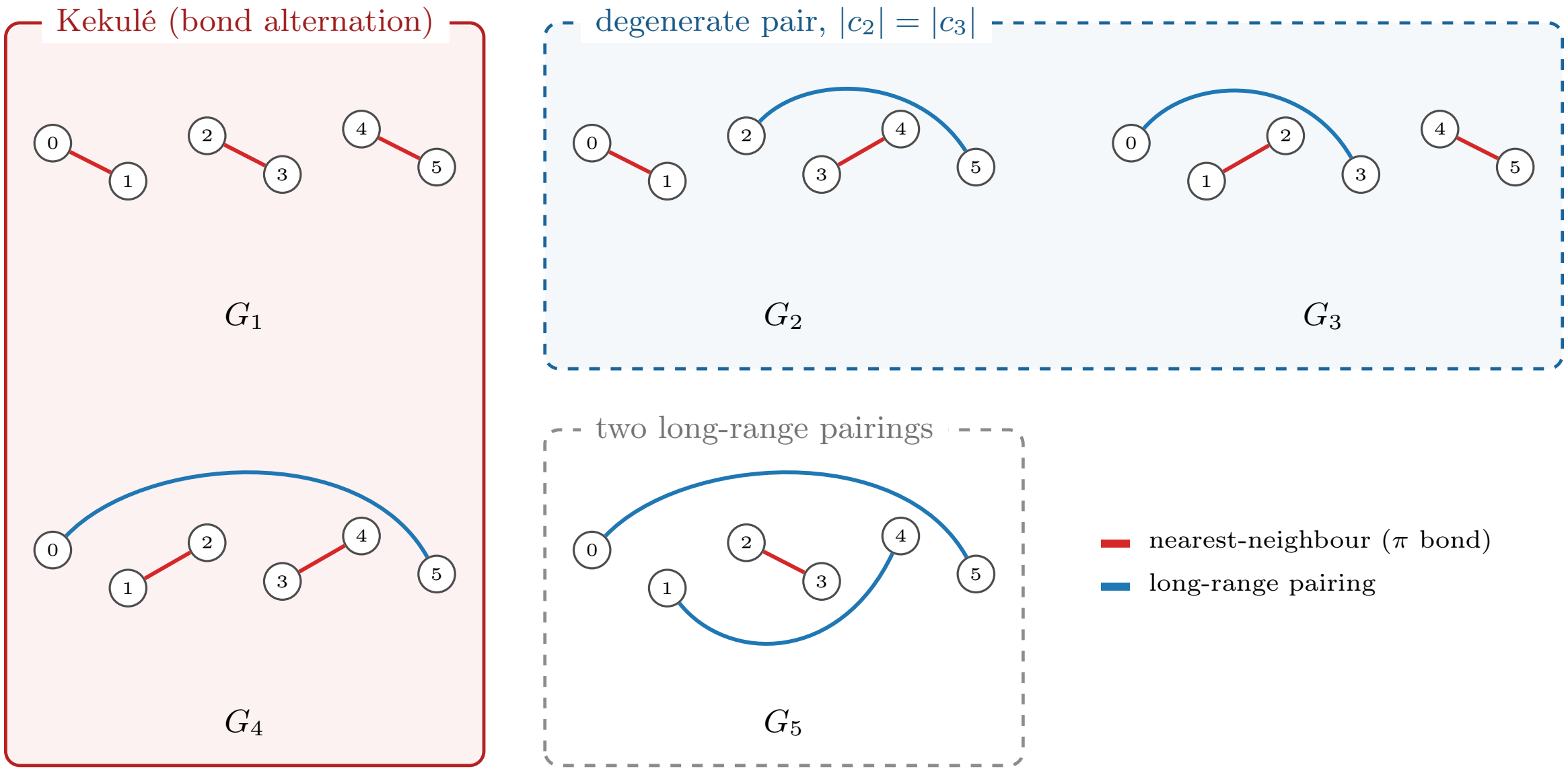
\begin{figure*}[t]
\centering
\begin{tikzpicture}[
  font=\tiny,
  cnode/.style={circle,draw=black!70,semithick,fill=white,
                inner sep=0.6pt,minimum size=3.6mm},
  nn/.style={draw=eA,line width=1.1pt},
  lr/.style={draw=eC,line width=1.1pt},
  gbox/.style={rounded corners=4pt,line width=0.9pt},
  gtag/.style={fill=white,inner xsep=4pt,inner ysep=1pt,font=\small},
]
% ---- one panel: #1 = shift, #2 = bonds, #3 = name/label ---------------
\newcommand{\hexpanel}[3]{%
  \begin{scope}[shift={(#1)},name prefix=#3-]
    \node[cnode] (c0) at (-1.887, 0.118) {0};
    \node[cnode] (c1) at (-1.147,-0.256) {1};
    \node[cnode] (c2) at (-0.370, 0.190) {2};
    \node[cnode] (c3) at ( 0.370,-0.190) {3};
    \node[cnode] (c4) at ( 1.147, 0.256) {4};
    \node[cnode] (c5) at ( 1.887,-0.118) {5};
    #2
  \end{scope}
  \node[font=\small] at ($(#1)+(0,-1.58)$) {$#3$};}
% ---------------- grouping boxes (drawn first, so fills sit behind) -----
\draw[gbox,draw=eA!85!black,fill=eA!6]  (-2.35,-6.00) rectangle (2.35,1.30);
\draw[gbox,draw=eC!85!black,dashed,fill=eC!5] (2.95,-2.10) rectangle (12.95,1.30);
\draw[gbox,draw=black!45,dashed] (2.95,-6.00) rectangle (7.65,-2.70);
% ---------------- row 1 ------------------------------------------------
\hexpanel{0,0}{\draw[nn] (c0)--(c1); \draw[nn] (c2)--(c3); \draw[nn] (c4)--(c5);}{G_1}
\hexpanel{5.3,0}{\draw[nn] (c0)--(c1); \draw[nn] (c3)--(c4);
  \draw[lr] (c2) .. controls (0.20,0.80) and (1.30,0.80) .. (c5);}{G_2}
\hexpanel{10.6,0}{\draw[nn] (c1)--(c2); \draw[nn] (c4)--(c5);
  \draw[lr] (c0) .. controls (-1.30,0.80) and (-0.20,0.80) .. (c3);}{G_3}
% ---------------- row 2 ------------------------------------------------
\hexpanel{0,-4.0}{\draw[nn] (c1)--(c2); \draw[nn] (c3)--(c4);
  \draw[lr] (c0) .. controls (-1.00,1.05) and (1.15,1.20) .. (c5);}{G_4}
\hexpanel{5.3,-4.0}{\draw[nn] (c2)--(c3);
  \draw[lr] (c0) .. controls (-1.00,1.05) and (1.15,1.20) .. (c5);
  \draw[lr] (c1) .. controls (-0.55,-1.00) and (0.55,-1.00) .. (c4);}{G_5}
% ---------------- box tags (on top, white fill breaks the border) -------
\node[gtag,text=eA!75!black,anchor=west] at (-2.00,1.30)
  {Kekul\'e (bond alternation)};
\node[gtag,text=eC!75!black,anchor=west] at (3.30,1.30)
  {degenerate pair,\ $|c_2|=|c_3|$};
\node[gtag,text=black!55,anchor=west] at (3.30,-2.70)
  {two long-range pairings};
% ---------------- legend ------------------------------------------------
\node[anchor=west,align=left,font=\scriptsize] at (8.30,-4.00)
  {\textcolor{eA}{\rule{8pt}{2.2pt}}~~nearest-neighbour ($\pi$ bond)\\[4pt]
   \textcolor{eC}{\rule{8pt}{2.2pt}}~~long-range pairing};
\end{tikzpicture}
\caption{The five valence bond structures used for the hexatriene $\pi$ system.
Vertices are the carbon $2p_z$ orbitals at their positions in the molecular
geometry, numbered $0$--$5$ along the chain, so the zig-zag of the backbone is
reproduced. Each structure is a perfect matching: every orbital carries one
electron pair and contributes one angle $\theta_e$. They are the non-crossing
(Rumer) matchings, the complete set of linearly independent singlet structures,
$5$ of the $15$ possible pairings. $G_1$ and $G_4$ are the two bond-alternation
(Kekul\'e) patterns; in a cyclic system such as benzene the two are related by
rotation and therefore exactly degenerate. The chain is open, however, so the
alternation of $G_4$ cannot close and its remaining pair $(0,5)$ spans the two
terminal carbons as a long bond. Only $G_1$ has every pair on an actual $\pi$
double bond, and it carries the dominant weight. $G_2$ and $G_3$ are images of
one another under the reflection $i\mapsto5-i$ of the chain and are therefore
exactly degenerate. $G_5$ is invariant under that reflection and pairs two sets
of non-bonded orbitals.}
\label{fig:hexgraphs}
\end{figure*}

\subsection{\label{limit}Limits of the Hydrogenic Representation}

The projection protocol presented in this work relies on the assumption that the
local electronic structure of a target molecule can be faithfully represented by
a collection of hydrogenic subsystems. While this approximation proved to be
remarkably accurate for the hydrocarbon systems investigated above, its
applicability is naturally limited by the underlying chemical bonding. 
We discuss three regimes in which this assumption is
strained: ionic dissociation and lone pairs \newline

\paragraph*{Ionic bonding.}
An illustrative counterexample is the dissociation of lithium hydride. Within the
present framework, the Li--H bond is represented by a hydrogenic two-center
system, which inherently favours a homolytic dissociation (see the $H_2$ dissociation example in Ref.~\cite{kottmann2023Molecular}) 
\[
\mathrm{LiH}\rightarrow\mathrm{Li}+\mathrm{H},
\]
that is also directly built into our applied predictor model via the boundary conditions. The observed dissociation of LiH is however heterolytic,
\[
\mathrm{LiH}\rightarrow\mathrm{Li}^{+}+\mathrm{H}^{-}.
\]
Consequently, the optimal hydrogenic projection cannot reproduce the electronic
structure along the dissociation coordinate. Even after optimizing the projection
radius over interatomic distances ranging from $0.5\,\text{\AA}$ to
$4.5\,\text{\AA}$, the best attainable approximation exhibits a mean energy error
of $13.61\,\mathrm{mE}_h$, with an order of magnitude above chemical accuracy and
than the largest hydrocarbon error reported in
\autoref{table:overview}.
Using a higher order mapping for the LiH
distance $d$ onto the hydrogenic distance
\begin{equation}
  R(d)=\sqrt{R_0^{2}+\big(s\,d+c\,d^{2})^{2}},
  \label{eq:lih_map}
\end{equation}
with three parameters fitted once to the SPA angles along the curve, the mean energy error drops to $ 0.078 \,\mathrm{mE}_h$.
Heterolytic bonds therefore are accessible to the hydrogenic framework, at
the cost of a bond-specific, nonlinear projection.

\paragraph*{Lone pairs.}
A second limitation is more constructive, in that it can be absorbed into the
existing framework at the price of a single additional fitted quantity. The
projection of \autoref{sec:projection} represents bonds: every SPA pair is
associated with an edge of the molecular graph and is mapped onto a two-center
hydrogenic system aligned with that edge. A lone pair (for example the two lone-pairs on the oxygen in $H_2O$) is also a localized
electron pair, but it is not shared between two nuclei and therefore has no edge
onto which it could be projected. Discarding it is equally unsatisfactory, since
the lone-pair density forms part of the local environment in which the
neighbouring bonds reside, and omitting it would for example misrepresent the couplings of
the O--H pairs in water.

\begin{figure}[t]
\centering
\begin{tikzpicture}[
  scale=1.05,
  Onode/.style={circle,fill=black,inner sep=1.7pt},
  Hreal/.style={circle,draw=black!75,fill=white,line width=0.6pt,inner sep=1.3pt},
  Hsur/.style={circle,draw=surH,fill=surH!20,line width=0.6pt,inner sep=1.3pt},
  pairline/.style={draw=surH,line width=1.1pt},
  font=\small]

  % projection sphere
  \draw[dashed,black!45] (0,0) circle (1.25);

  % oxygen
  \node[Onode] (O) at (0,0) {};
  \node[font=\scriptsize,anchor=west] at (0.10,-0.16) {O};

  % O--H bonds: surrogate connector on the sphere, real H outside
  \foreach \a in {218,322}{
    \draw[black!35] (0,0) -- (\a:2.05);
    \draw[pairline] (\a:1.25) -- (\a:2.05);
    \node[Hsur]  at (\a:1.25) {};
    \node[Hreal] at (\a:2.05) {};
  }
  \node[font=\scriptsize,anchor=north] at (322:2.28) {H};
  \node[font=\scriptsize,anchor=north] at (218:2.28) {H};

  % lone pairs: surrogate hydrogen dimers
  \foreach \a in {35,145}{
    \draw[pairline] (\a:1.25) -- (\a:1.88);
    \node[Hsur] at (\a:1.25) {};
    \node[Hsur] at (\a:1.88) {};
  }

  % annotations
  \draw[<->,black!70,line width=0.5pt] (0,0) -- (180:1.25)
        node[midway,above,font=\scriptsize] {$r$};
  \node[font=\scriptsize,anchor=east] at (145:1.60) {$d$};
\end{tikzpicture}
\caption{Hydrogenic projection of water including lone pairs. The two O--H bonds
are projected as before: a surrogate hydrogen (blue) is placed on the sphere of
radius $r$ centered at the oxygen and paired with the real hydrogen (white).
Each of the two lone pairs is represented by a surrogate hydrogen dimer along the
lone-pair direction, with the inner atom on the same sphere of radius $r$ and the
outer atom displaced by an additional distance $d$, the only new parameter. The
valence structure of water is thereby mapped onto an $\mathrm{H}_8$ system.}
\label{fig:water_projection}
\end{figure}
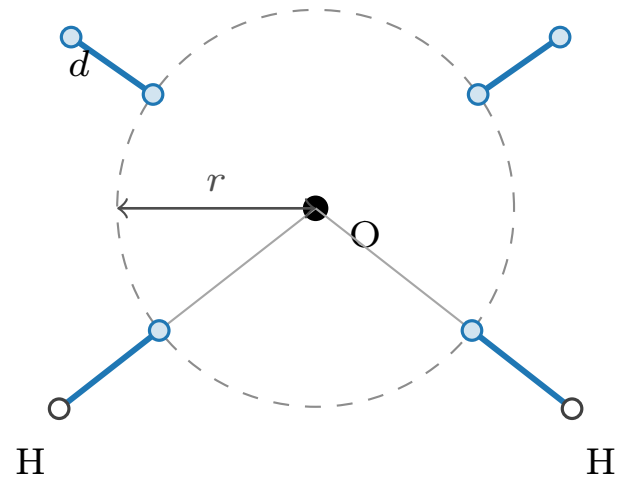

Our initial fix is to represent each lone pair by a surrogate hydrogen dimer
oriented along the lone-pair direction, as illustrated for water in
\autoref{fig:water_projection}, without promoting it to an additional variational
degree of freedom. The angles associated with the surrogate dimers are held fixed
at
\begin{equation}
\label{eq:lp_angle}
    \theta_{\mathrm{lp}} = 0 ,
\end{equation}
so that these pairs remain in their reference configuration and contribute no
parameter of their own. Their function is solely to be present: the
additional centers enter the Hamiltonian of the hydrogenic system and thereby
modify the effective couplings seen by the genuine O--H edges. The lone pairs act
as spectators that shape the environment of the bonds, rather than as pairs whose
own correlation is transferred.

What has to be parametrized is consequently not an angle but a placement. The
inner surrogate atom is fixed on the same projection sphere of radius $r$ that
carries the bond connectors, so that lone pairs and bonds are treated on an equal
footing, while the outer atom is displaced radially by a distance $d$. This
distance sets how strongly the surrogate dimer perturbs its surroundings, and is
fitted once per lone-pair type such that the combined influence of all
surrogate dimers reproduces the reference O--H parameters. The fit is necessarily
a joint one: in water both lone pairs act on the same two O--H pairs
simultaneously, so $d$ must be chosen so that their superposed effect, not that
of an isolated dimer, is correct.

For water this yields four surrogate hydrogen atoms in addition to the two
projected O--H connectors, mapping the valence structure onto an $\mathrm{H}_8$
system of the same size as the methane projection of
\autoref{fig:projection_methane}, but with only two of its four pairs carrying
transferred angles, the remaining two being fixed by \autoref{eq:lp_angle}. The
construction has a mean transfer error of $0.82\,\mathrm{mE}_h$,
still below chemical accuracy, although the largest compared to the hydrocarbons. The cost of the extension is therefore one
additional fitted distance per lone-pair environment, while the number of
transferred variational parameters remains equal to the number of genuine bonds.

This \textit{ad-hoc} fix seems promising for the near future, potential extensions could introduce Helium atoms in combination with chargeless ghost to produce more complicated but ideally also more accurate models. The price is that the underlying machine learning model needs to be trained on 3 types of atoms instead of just one.\\

Possible routes for extending the framework further, with a more flexible
projection or a data-driven parameter model, are discussed in
\autoref{sec:conclusion}.

\section{\label{sec:conclusion}Conclusion and Outlook}

We started from a premise rather than a construction: that the variational
parameters of small hydrogenic systems can be obtained cheaply, by whatever
means. What this work adds is the machinery that makes such a black box useful
for molecules that are not hydrogenic, namely a geometric projection of the
local bonding environment onto hydrogenic fragments, together with a transfer
rule that reassembles the returned parameters into the circuit of the full
system. Solver and circuit family enter only as interfaces; the projection
carries the physics, and it is fitted once, with a single radius per bond type,
and thereafter reused unchanged.

The protocol was calibrated on methane and ethane and evaluated on propane and
norbornane, neither of which required refitting. Transfer errors stayed below
chemical accuracy throughout, from $0.022\,\mathrm{mE}_h$ for methane to
$0.553\,\mathrm{mE}_h$ for the strained bicyclic norbornane, and the C--H and
C--C contributions were found to combine additively rather than
multiplicatively, so that the error grows only linearly with the number of
projected bonds. Used as initial states rather than as final answers, the
transferred parameters roughly halved the number of circuit evaluations required
for convergence; the accuracy with our cheap prediction model is, however, already quite good and might already be sufficient for most applications.

That the ansatz is a slot rather than an ingredient is most visible in the
conjugated systems. The projection fitted on saturated hydrocarbons was carried
over unchanged to a multiconfigurational valence-bond wavefunction for butadiene
and hexatriene, where it supplies the angles of every valence bond structure
entering the expansion. Up to four structures, the transfer costs at most
$0.03\,\mathrm{mE}_h$, below $0.07\,\%$ of the total error, so that the
projected and the directly optimized results are indistinguishable while the
ansatz itself becomes markedly more accurate. Only the most delocalized
structure, carrying two long-range pairings, escapes the transfer, which is
consistent with the locality the projection rests on and illustrates a limit of the current method.

Further limitations and potential fixes were identified: Lithium hydride, whose
dissociation is ionic rather than homolytic, is misrepresented by a covalent
two-center model by $13.61\,\mathrm{mE}_h$. Lone pairs, by contrast, can be
absorbed by adding surrogate centers that carry no parameters of their own, at
the cost of one further fitted distance and a residual error of
$0.82\,\mathrm{mE}_h$ for water. Genuine $\pi$ bonding lies outside a purely
radial projection altogether. The protocol is therefore best read as a
physically motivated approximation whose accuracy is set by how faithfully a
hydrogenic system can stand in for the local environment.

Two directions follow. The projection can be made more flexible, admitting
higher-order geometric transformations or surrogate centers placed off the bond
axis, so that asymmetric and partially ionic environments come within reach.
Alternatively, the analytical parameter model can be replaced by a data-driven
one~\cite{vpm}, for instance a graph neural network learning parameters directly
from the local molecular environment. Since solver, projection and ansatz remain
separable, progress on any one of them propagates to the others without altering
the protocol, which is what would establish parameter transfer as a general
route to scalable initial-state preparation in quantum chemistry.

%------------------------------------
\newpage
\onecolumngrid
\appendix
 
\section{Appendixes}
% \subsection{\label{sec:spa_example}SPA Implementation}

% \begin{lstlisting}[style=minimalpython]
% """ SPA Water example with tequila. """
% geometry = "O  0.0  0.0  0.117\n 
%             H  0.0  0.757  -0.469 \n 
%             H  0.0  -0.757  -0.469 \n"
            
% mol = tq.Molecule(geometry=geometry, basis_set="sto-3g")

% mol, graph = sun.CLPO.generate_CLPO_molecule_edges(mol)

% E = tq.ExpectationValue(
%     U=U,
%     H=mol.make_hardcore_boson_hamiltonian())

% result = tq.minimize(E).energy

% \end{lstlisting}

\subsection{\label{sec:initial}Initial State Preparation of Hydrogenic Systems}

To evaluate the effectiveness of the proposed initial-state preparation strategy, we consider a series of linear hydrogenic systems with interatomic separations ranging from \(0.5~\text{\AA}\) to \(3.0~\text{\AA}\). Small hydrogenic systems constitute the fundamental building blocks of the hierarchical decomposition scheme introduced in this work and therefore provide a suitable benchmark for assessing the quality of the predicted SPA parameters.

\begin{table}[H]
\centering
\small
\setlength{\tabcolsep}{9pt}
\begin{tabular}{lccccc}
\toprule
\multirow{3}{*}{\textbf{Molecule}} & \multirow{3}{*}{\textbf{Error (mEh)}}
  & \multicolumn{4}{c}{\textbf{Iterations}} \\
\cmidrule(l){3-6}
 & & \multicolumn{2}{c}{2-point} & \multicolumn{2}{c}{analytical} \\
\cmidrule(lr){3-4}\cmidrule(l){5-6}
 & & $\tilde{\theta}$ & $0$ & $\tilde{\theta}$ & $0$ \\
\midrule
H$_2$     & 0.00 & --  &  8 & -- & 4 \\
H$_4$     & 0.26 &  7  & 13 &  2 & 4 \\
H$_6$     & 1.01 & 11  & 20 &  3 & 5 \\
H$_8$     & 1.75 & 13  & 25 &  3 & 5 \\
H$_{10}$  & 2.49 & 16  & 30 &  3 & 5 \\
H$_{12}$  & 3.24 & 19  & 35 &  3 & 5 \\
\bottomrule
\end{tabular}
\caption{Performance of the parameter modelling method on linear hydrogenic
systems. The error denotes the energy difference
$|E[\psi(\theta^\star)]-E[\psi(\tilde{\theta})]|$ between the simulated
expectation values of states generated from SPA circuits with fully optimized
parameters $\theta^\star$ and with the parameters $\tilde{\theta}$ predicted by
the read-out scheme of the main text. The iteration counts are the average
number of circuit evaluations the SPA optimization needs to converge; the two
sub-columns give the initialization, either the default $\theta=0$ or the warm
start $\theta=\tilde{\theta}$, and the two groups the gradient used, a numerical
2-point finite difference or the analytical gradient.}
\label{table:diag}
\end{table}

\section{Data access}
The code for the experiments is available in the repository: \url{https://github.com/tequilahub/AlchemicalTransferability}.

\newpage
\twocolumngrid
\nocite{*}
\bibliography{paper}

\end{document}